# Non equilibrium velocity distributions of $H^-$ ions in $H_2$ plasmas and photodetachment measurements

P.Diomede[1,*], S.Longo[1,2] and M.Capitelli[1,2]

[1] *Dipartimento di Chimica dell'Università di Bari, Via Orabona 4, 70126 Bari, Italy*

[2] *IMIP/CNR, Via Orabona 4, 70126 Bari, Italy*

* e-mail:diomede@chimica.uniba.it

## ABSTRACT

A theoretical study of the energy distribution function of the negative hydrogen ion $H^-$ in typical conditions for multicusp ion sources is presented. The case of $H/H_2$ mixture is studied by a Monte Carlo solution of the kinetic equation for $H^-$ transport. A simple analytical model is obtained for the case of a fully dissociated H plasma and uniform reduced electric field (E/N). Results are in good agreement with the two-group distribution deduced from laser photodetachment experiments and explain the low energy group as an effect of the charge exchange collisions of negative ions with H atoms.

**Introduction**

Hydrogen plasmas are proposed as sources for the production of neutral beams generated by neutralization of high-energy negative ion beams to be used in nuclear fusion experiments [1, 2]. The knowledge of the negative ion temperature is thus very important and much effort has been done, both from the theoretical and from the experimental point of view, in order to explain the mechanisms that determine its peculiar distribution.

In a recent paper by Ivanov [3] the existence of two negative ion groups with different temperatures in volume H⁻ sources is stated, starting from a bi-Maxwellian fit of photodetachment experimental data. The origin of these two populations is attributed to the existence of two H⁻ production regions with different plasma potentials: the low temperature corresponds to the ions formed in the central source region, while the higher temperature population is characterized by ions produced in the driver region and accelerated by the plasma potential into the central plasma region.

In the work by Mizuno et al. [4], concerning a one dimensional analysis, by means of a particle in cell simulation, of the effect of the ambipolar potential on the H⁻ ion density recovery after a laser photodetachment, it is shown that the negative ion temperature can be influenced by the ambipolar potential especially for the cases in which the negative ion density is comparable with the positive ion one.

In this work, using realistic cross sections and with no additional hypothesis, a test particle Monte Carlo simulation for the dynamics of H⁻ ions in $H_2$ plasmas is performed in the frame of a uniform reduced electric field for the same pressure of the experimental conditions of ref. [5]. The simulation is performed for different values of the dissociation degree. A simple analytical model is obtained for the case of a fully dissociated hydrogen plasma and uniform reduced electric field (E/N).

The H⁻ ion energy distribution function (iedf) for a dissociation degree similar to the experimental conditions of ref. [5] is compared with the ones obtained by fitting experimental data in the case of the one-temperature (ballistic kinetic theory) and two-temperature approximation.

Results are in good agreement with the two-group distribution deduced from laser photodetachment experiments and explain the low energy group as an effect of the charge exchange collisions of negative ions with H atoms.

In a recent paper by Hatayama et al. [6] a non equilibrium iedf is calculated by taking into account the space variation of the electric potential, but neglecting ion-neutral collisions. The authors found that the low energy part of the iedf as deduced from photodetachment experiments could not be justified from the model results. In the light of the present work, this is correct since the inclusion of charge exchange collisions with H atoms is essential to capture the low energy component.

**Monte Carlo simulation**

We assume a uniform plasma model, where the iedf is supposed to result from the collisional kinetics of negative ions in the weakly ionized plasma with neutral particles, here H atoms and $H_2$ molecules, and solve the transport equation for negative ions using realistic cross sections to include the following two collision processes

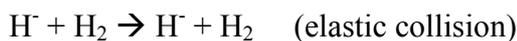
H⁻ + $H_2$ → H⁻ + $H_2$    (elastic collision)

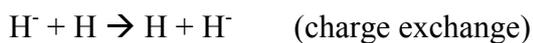
H⁻ + H → H + H⁻    (charge exchange)

Since in this case the ion mass is comparable with that of neutral particles, the Fokker-Planck treatments associated with either the Rayleigh or Lorentz limits cannot be applied and the linear Boltzmann equation for dilute ion transport must be solved to find the unknown steady-state negative ion velocity distribution $f(\mathbf{v})$. In the specific case the Boltzmann equation is written in the following form:

$$q\mathbf{E}/m \cdot \frac{\partial f(\mathbf{v})}{\partial \mathbf{v}} =$$
$$\int |\mathbf{v} - \mathbf{v}_{H_2}| \sigma_{el}(|\mathbf{v} - \mathbf{v}_{H_2}|, \vartheta)(f'F'_{H_2} - fF_{H_2}) \sin\vartheta d\vartheta d\varphi d^3v_{H_2} \quad (1)$$
$$+ (F_H - f) \int |\mathbf{v} - \mathbf{v}_H| \sigma_{ce}(|\mathbf{v} - \mathbf{v}_H|) F_H d^3v_H$$

where $\sigma_{el}$ and $\sigma_{ce}$ are the differential elastic and the total resonant charge exchange cross sections and F is a Maxwell distribution at the gas temperature.

As shown in ref. [7] such a solution can be obtained by a null collision test particle Monte Carlo method that takes into account the thermal distribution of the target particles.

Regarding the elastic collisions with the hydrogen molecule they are treated assuming isotropic scattering in the center-of-mass frame and introducing the momentum transfer cross section of ref. [8]. The resonant charge exchange with the hydrogen atom is treated without restrictions and using the cross sections from ref. [9]. The available data have been extended to cover the whole requested energy range by piecewise linear interpolation and extrapolation. Because of the low density of data at low energy in the case of the charge exchange process the results could be slightly different from those reported here if different interpolation techniques are used.

The method of ref. [7] has been extended in order to calculate the collision frequency for a species colliding with two different neutral targets, in this case the $H_2$ and the H. The collision probabilities for the two collision processes, for a given H- velocity are :

$$p_{mt} = \frac{n_{H_2}\sigma_{mt}(|\mathbf{v}-\mathbf{v}_{H_2}|)|\mathbf{v}-\mathbf{v}_{H_2}|}{v_{max}} \qquad (2)$$

$$p_{ce} = \frac{n_{H}\sigma_{ce}(|\mathbf{v}-\mathbf{v}_{H}|)|\mathbf{v}-\mathbf{v}_{H}|}{v_{max}} \qquad (3)$$

$$v_{max} = \max_{v, v_H, v_{H_2}} (n_{H_2}\sigma_{mt}(|\mathbf{v}-\mathbf{v}_{H_2}|)|\mathbf{v}-\mathbf{v}_{H_2}| + n_{H}\sigma_{ce}(|\mathbf{v}-\mathbf{v}_{H}|)|\mathbf{v}-\mathbf{v}_{H}|) \qquad (4)$$

where $n_{H_2}$ and $n_H$ are, respectively, the $H_2$ and the H number densities, $\sigma_{mt}$ is the total momentum transfer cross section.

The velocity of the target particle is selected by means of the von Neumann rejection technique from a Maxwell distribution of neutral particles with a temperature of 300K.

The simulation has been performed for a gas pressure of 3mtorr, the density for the H atom is $9\times10^{19} m^{-3}$ (the same as in the experimental conditions as reported in ref. [5]). In fig.1 the calculated H$^-$ iedf for different values of the reduced electric field, calculated taking into account the density of both the H atom and the $H_2$ molecule, is shown. The non equilibrium features can be easily inferred, specially as regards the higher values of the reduced electric field.

In fig. 2, the iedf obtained for a fully dissociated gas and different values of the reduced electric field is displayed; in this case the shape is determined by the effect of the charge exchange collisions and the electric field and a lower energy content can be noted with respect to the curves in fig. 1 corresponding to the same value of the reduced electric field.

The case of a non dissociated gas, in which the H$^-$ ions undergo only elastic collisions is illustrated in fig. 3: here the superthermal component of the iedf is much less conspicuous, suggesting that H$^-$-H collisions are essential to observe this feature. This point is further elaborated in the next section.

**Approximate theory for H plasma**

The results of the previous section suggest that the low energy component of the iedf is essentially due to charge exchange collisions of negative ions with the atomic fraction. Under such conditions the only important collision process is charge exchange H/H⁻, i.e. any H⁻ ion basically starts from a Maxwellian distribution at temperature T and constantly accelerates under the force –e**E** until a collision leads it back to thermal energy. Since kT is much smaller than the average ion energy we can assume that the post collision velocity is negligible and the iedf becomes the product of two factors: the mean residence time of the ion in the energy interval εdε given by (dt/dε) dε and an absorption factor which represents the collision process, i.e.

$$f(\varepsilon) \propto \left(\frac{d\varepsilon}{dt}\right)^{-1} \exp\left(-\int \frac{1}{\lambda(\varepsilon)}dx\right), \qquad (5)$$

where $\lambda(\varepsilon)$ is the instantaneous mean free path.

By expressing all quantities in terms of the kinetic energy the following expression is simply found:

$$f(\varepsilon) \propto \frac{1}{\sqrt{\varepsilon}} \cdot \exp\left(-\frac{1}{eE/n_H} \int_0^\varepsilon \sigma_{ce}(\xi)d\xi\right) \qquad (6)$$

If we further assume that the cross section is constant in the energy range considered the result is a Maxwell distribution at $T_{high} = eE/k\sigma n_H$ modulated by a factor $1/\varepsilon$ which produces a low energy component: under this light the low energy group in the iedf is explained as a result of the kinematics of ion acceleration. This effect is clearly shown by the theoretical plot in fig. 4 for different E/N values: the curves are quite the same assuming either a realistic cross section or a constant cross section equal to its average value in the energy range considered. A very good agreement with the Monte Carlo calculations results is obtained.

**Interpretation of photodetachment results**

Following Ivanov [3] it is possible to deduce a bimodal iedf from photodetachment data in the framework of the ballistic kinetic theory (BKT) namely the assumption of collisionless and inertial filling of the negative ion space hole produced by the laser. Unfortunately the calculation of this reference are somewhat difficult to reproduce since the laser beam radius is not reported, therefore we apply the method to a different series of data.

According to ref. [5] the depleted region is represented as a cylinder with an aspect ratio R/h << 1. Under such conditions the ion density along the cylinder axis as a function of time is easily obtained by applying time reversal to ion trajectory converging to the measurement point, and the result is:

$$n^-(r=0,t) = n_0^- \iiint f_0(\mathbf{v}) H(v_r - R/t) d^3v \qquad (7)$$

Where $f_0$ is the ion velocity distribution function at the steady state, H is the Heaviside function, R is the laser spot radius, r is the radial distance from the cylinder axis, $v_r$ is the radial component of the ion velocity, $n_0^-$ is the steady state value of the negative ion number density, t is the time interval between the two laser shots.

This expression is evaluated straightforwardly in the case of a multi-modal Maxwellian distribution, i.e.

$$f_0(\mathbf{v}) = n_0^- \sum_{i=1}^{n} \chi_i (\beta_i / \pi)^{3/2} \exp(-\beta_i v^2), \qquad (8)$$

where $\beta_i = m/2kT_i$ and $\chi_1 + ... \chi_n = 1$, giving

$$n^-(t) = n_0^- \sum_i \chi_i \exp(-\beta_i R^2 / t^2). \qquad (9)$$

A least square fit of the two series of experimental data of ref. [5] concerning the recovery of the H$^-$ density at r = 0 for a laser spot of radius 2mm and 4mm respectively and for a multicusp negative ion discharge at 3mtorr, has been carried out simultaneously according to the following equations which are equivalent to (3) in the case n = 1 and n = 2 respectively:

$$n^- = n_0^- \exp[-R^2/(v_{th}t)^2] \qquad (10)$$

$$n^-/n_0^- = a \exp[-R^2/(v_{th}t)^2] + (1-a) \exp[-R^2/(bv_{th}t)^2] \qquad (11)$$

where the fit parameters are the thermal velocity of negative ions $v_{th} = \sqrt{2kT_-/m_-}$ in the first case and $a$, $b$ and $v_{th}$ in the second. In the first case $v_{th}$ = 6917m/s and the mean square error is equal to $8.45 \times 10^{-3}$. The values found for the second case are $a$ = 0.616, $b$ = 0.365 and $v_{th}$ = 10223m/s (with a mean square error equal to $3.86 \times 10^{-3}$) corresponding to a temperature $T_1$ = 6286K, while the temperature of the second Boltzmann distribution is $T_2 = b^2 T_1$ = 817K. The comparison of the experimental data with the analytical fits is displayed in figs. 5-6.

The values for average ion energy deduced in the two cases are respectively

$$3kT/2e = \qquad 0.37 \text{ eV} \qquad \text{for n = 1} \qquad (12)$$

$$3k(\chi_1 T_1 + \chi_2 T_2)/2e = \qquad 0.55 \text{ eV} \qquad \text{for n = 2} \qquad (13)$$

where $\chi_1 = a$, $\chi_2 = 1-a$.

A comparison between the iedf obtained by fitting the experimental data and the one obtained from the Monte Carlo simulation described previously is performed.

The reduced electric field E/N has been chosen in order to have the same average kinetic energy of the one and two temperature Boltzmann iedf i.e. 1300Td for the case n = 1 and 1800Td for the case n = 2. These values are calculated taking into account the number density of both the H and $H_2$ particles.

A quite good agreement between the two temperature iedf and the one resulting from the simulation can be noted (figs. 7-8). This agreement is not expected to be substantially changed by including inelastic and reactive processes because of the relative high threshold: e.g. the $H^-/H_2$ detachment process has a threshold of 2.37 eV [8]. This further result shows that a two-temperature approximation gives a good description of the real energy distribution of negative ions in the steady state and also allows to explain the appearance of an approximately bimodal distribution: the low temperature, almost thermal, component represents slow ions which have recently undergone a charge exchange collision with H atoms, while the superthermal "tail" represents a fraction of ions with higher "age" from the last collision with H and meanwhile accelerated by the field to a relative high energy while performing only elastic collisions with molecules.

**Conclusions**

We have studied the non equilibrium iedf of $H^-$ ion in cold $H_2$ plasmas in negative ion sources conditions under the simple physical hypothesis that the plasma is approximately uniform and that field acceleration and collision processes with neutral particles are essential, in opposition to previous studies where only Coulomb collisions are included. A Monte Carlo simulation including elastic collisions with $H_2$ molecules and charge exchange collisions with H, shows a pronounced bimodal behavior unless H is absent. A simple model of uniform acceleration and collisions applied to the H/H$^-$ case explains the presence of the low energy population fraction in the iedf as an effect of charge exchange processes with the atomic fraction. Results agree quite well with bimodal iedf obtained by inversion of experimental H$^-$ laser photodetachment data with the same dissociation degree and average kinetic energy (therefore E/N is not a free parameter).

These results can make the way to an improved procedure where inversion of the related integral operator is used to determine the iedf from experimental photodetachment data. Of course the quantitative description of the ion dynamics could be refined in the future by including space charge effect as in ref. [4] to further close the residual gap between theory and experiments, even if it will result in a nonlinear optimizations problem to be solved. Future studies should also consider the effect of the dissociation degree and the H atom temperature on the calculated iedf, and maybe include a superthermal H fraction as measured by Döbele et al [10].

**Acknowledgements**

The work was partially supported by MIUR (project no. 2005033911_002 and no. 2005039049_005).

**Figure captions**

**Fig. 1** H⁻ iedf calculated by means of the Monte Carlo technique for different values of the reduced electric field, for a gas pressure of 3 mtorr and an H atom number density of $9 \times 10^{19} m^{-3}$. The values of the reduced electric field, for the curves from 1 to 6 are , respectively, 500, 1000, 1500, 1800, 2100 and 2400Td.

**Fig. 2** H⁻ iedf calculated by means of the Monte Carlo technique for different values of the reduced electric field, for a gas pressure of 3 mtorr and a fully dissociated gas. The values of the reduced electric field, for the curves from 1 to 7 are , respectively, 500, 1000, 1500, 1800, 2100, 2400Td and 3600Td.

**Fig. 3** H⁻ iedf calculated by means of the Monte Carlo technique for different values of the reduced electric field, for a gas pressure of 3 mtorr and a non dissociated gas. The values of the reduced electric field, for the curves from 1 to 7 are , respectively, 20, 30, 40, 50, 60, 80, 100Td.

**Fig. 4** Comparison between the H⁻ iedf calculated by means of the Monte Carlo technique (dots), for a gas pressure of 3 mtorr and a fully dissociated gas, and the iedf obtained with the approximate theory for H plasma for different values of the reduced electric field (full lines). The values of the reduced electric field, from bottom to top, are respectively, 2100, 2400 and 3600Td.

**Fig. 5** Recovery of the H⁻ density at r = 0 for R = 0.2cm: experimental results (dots, ref [5]) least square fit with the one (dashed line) and two (full line) temperature ballistic approximation.

**Fig. 6** Recovery of the H⁻ density at r = 0 for R = 0.4cm: experimental results (dots, ref [5]) and least square fit with the one (dashed line) and two (full line) temperature ballistic approximation.

**Fig. 7** H⁻ ion energy distribution function (iedf) as a function of energy as obtained from a Boltzmann distribution with the temperature of the analytical fit (full line) and a test particle Monte Carlo simulation (dotted line). The two distributions are characterized by the same average kinetic energy.

**Fig. 8** H⁻ ion energy distribution function (iedf) as a function of energy as obtained from a two temperature Boltzmann distribution with the parameters of the analytical fit (full line) and a test particle Monte Carlo simulation (dotted line). The two distributions are characterized by the same average kinetic energy.

Fig. 1

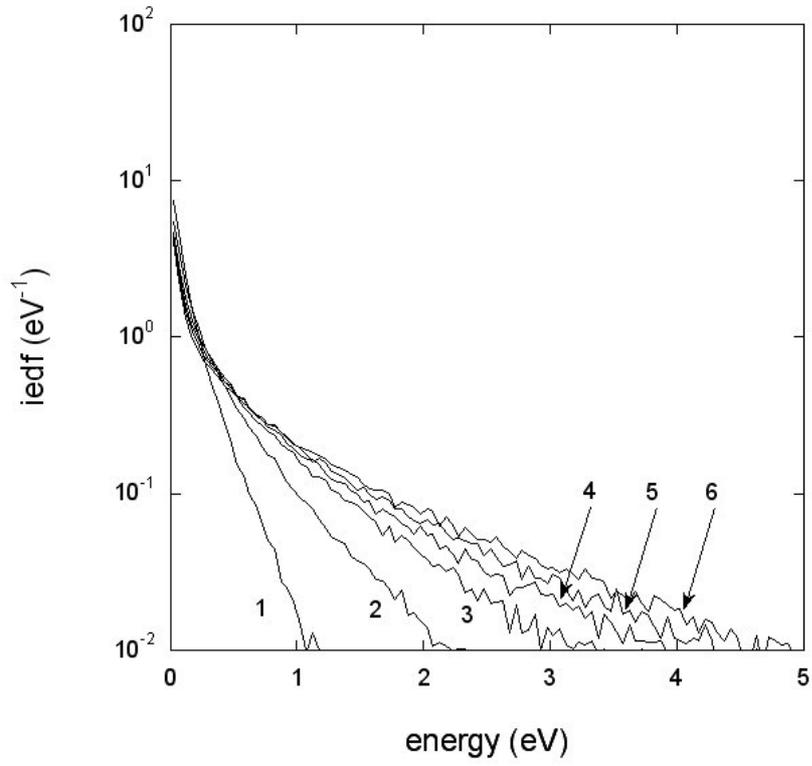

Fig. 2

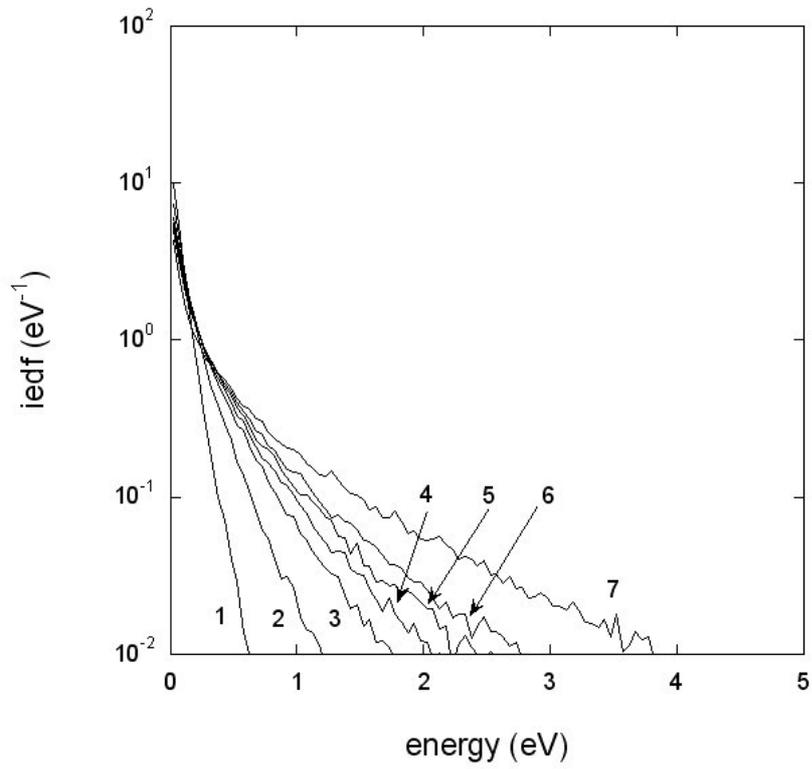

Fig. 3

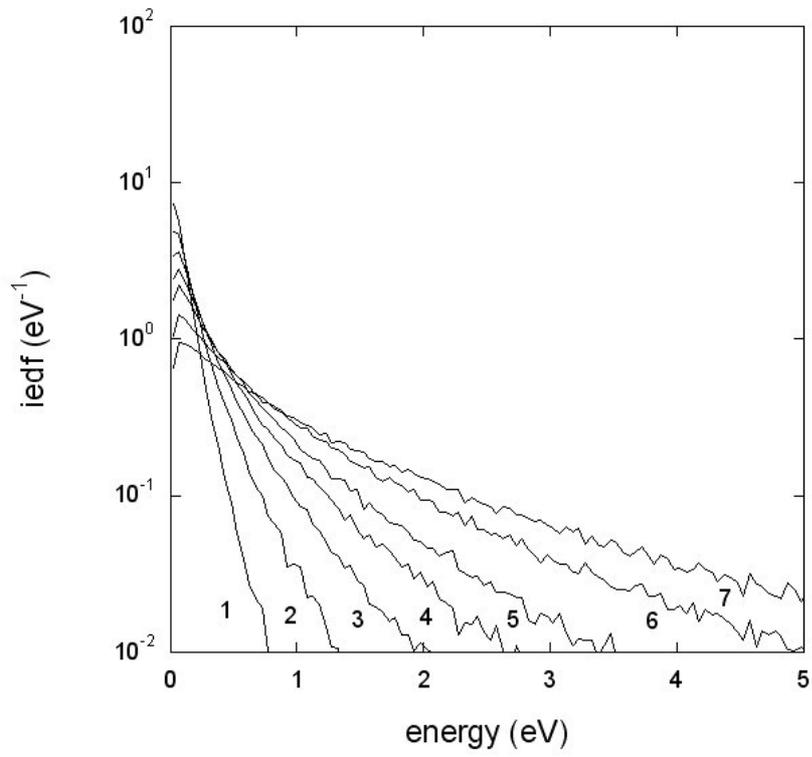

Fig. 4

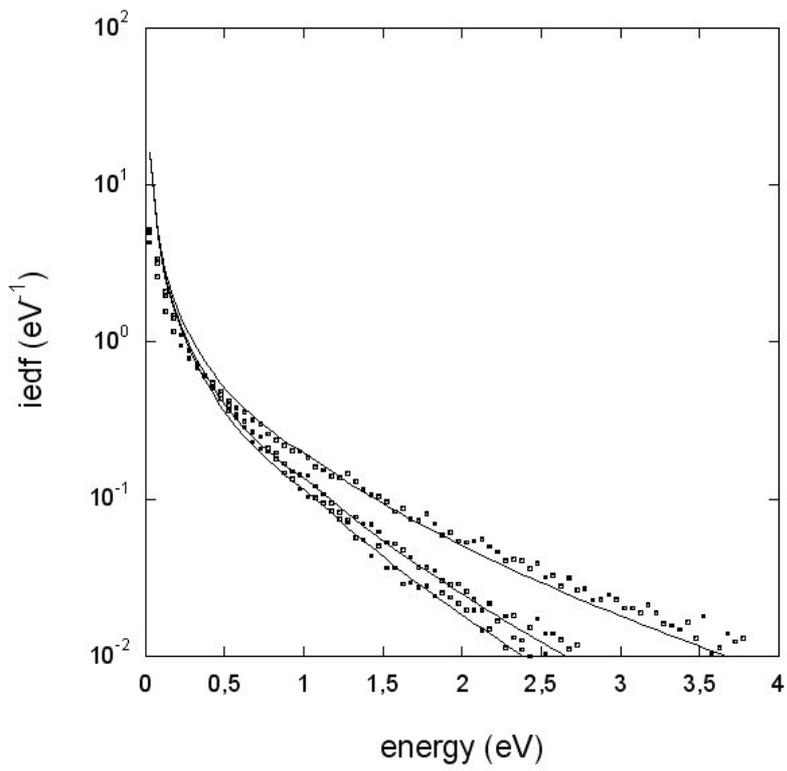

Fig.5

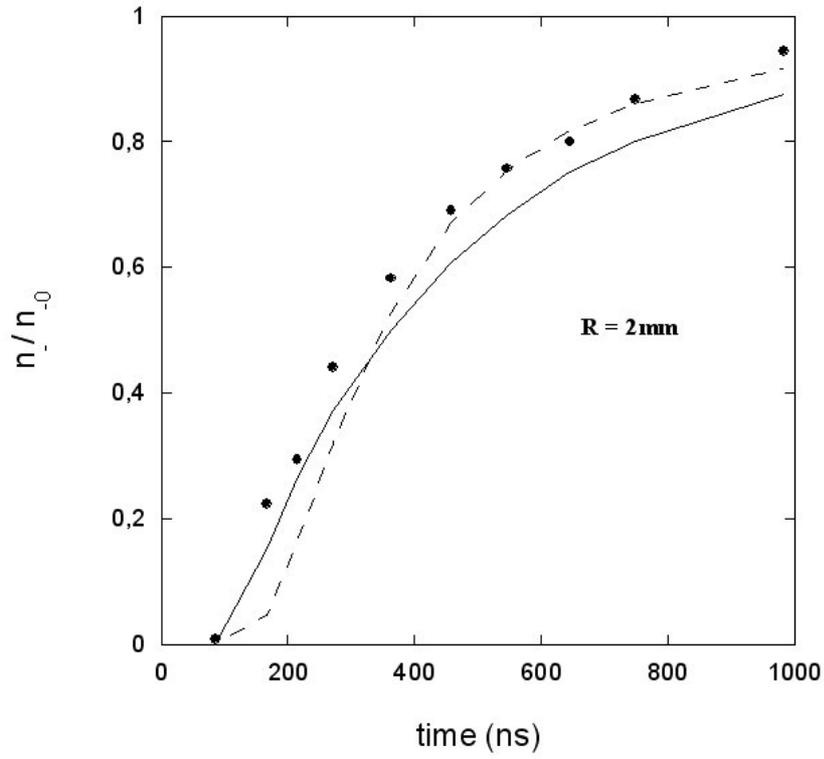

Fig. 6

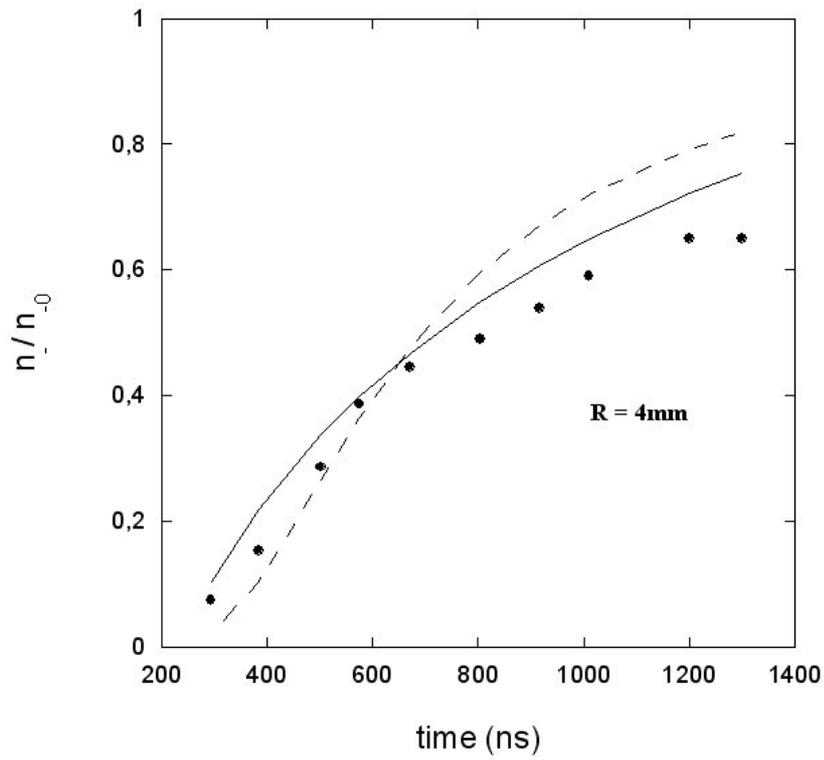

Fig. 7

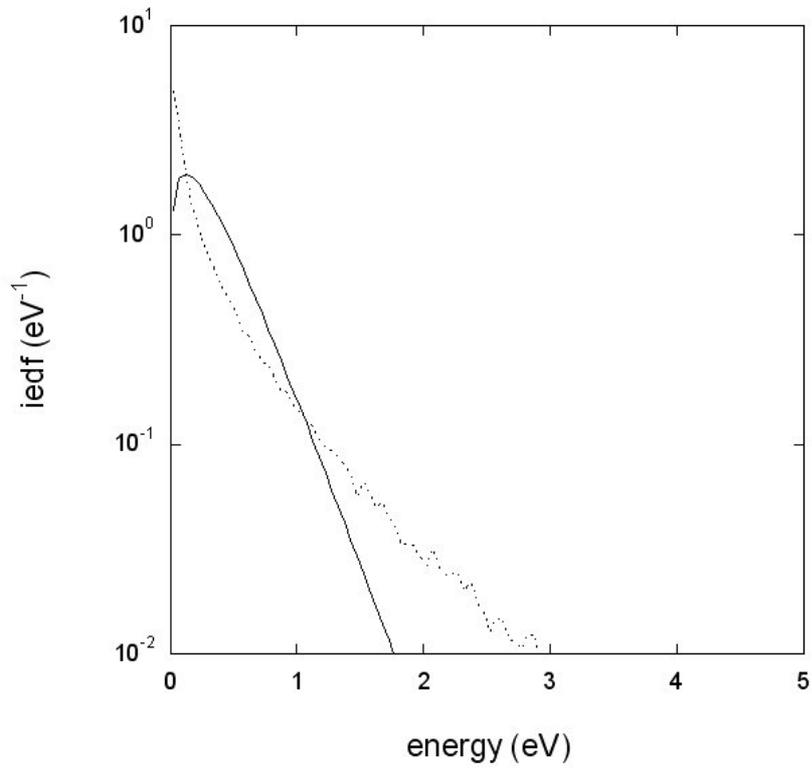

Fig. 8

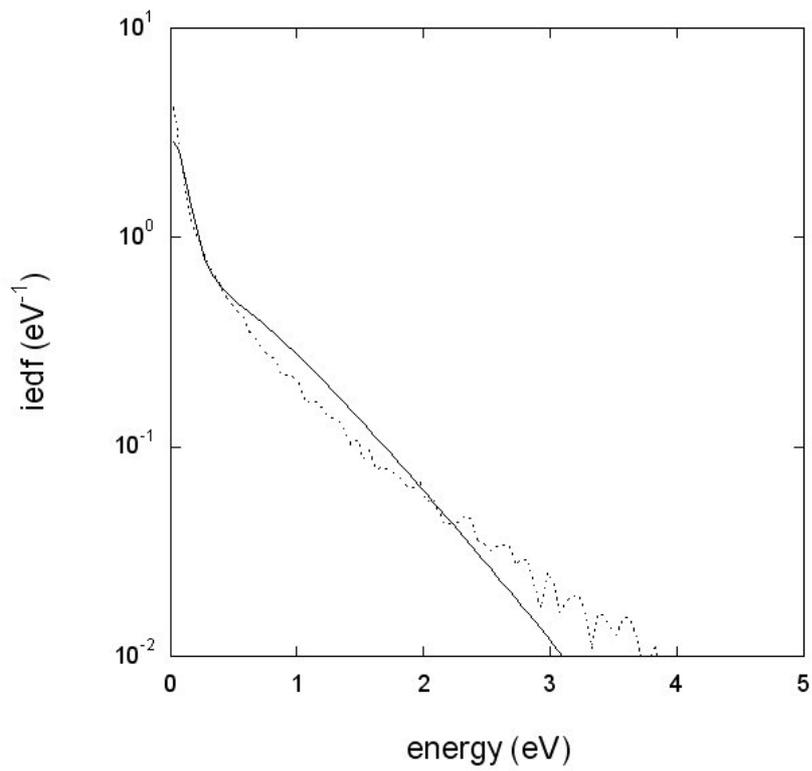